**From What to How: An Initial Review of Publicly Available AI Ethics Tools, Methods and Research to Translate Principles into Practices**

Jessica Morley[1], Luciano Floridi[1,2], Libby Kinsey[3], Anat Elhalal[3]

[1] Oxford Internet Institute, University of Oxford, 1 St Giles', Oxford, OX1 3JS
[2] Alan Turing Institute, British Library, 96 Euston Rd, London NW1 2DB
[3] Digital Catapult, 101 Euston Road, Kings Cross, London, NW1 2RA

**Abstract**

The debate about the ethical implications of Artificial Intelligence dates from the 1960s (Samuel, 1960; Wiener, 1961). However, in recent years symbolic AI has been complemented and sometimes replaced by (Deep) Neural Networks and Machine Learning (ML) techniques. This has vastly increased its potential utility and impact on society, with the consequence that the ethical debate has gone mainstream. Such a debate has primarily focused on principles—the 'what' of AI ethics (beneficence, non-maleficence, autonomy, justice and explicability)—rather than on practices, the 'how.' Awareness of the potential issues is increasing at a fast rate, but the AI community's ability to take action to mitigate the associated risks is still at its infancy. Our intention in presenting this research is to contribute to closing the gap between principles and practices by constructing a typology that may help practically-minded developers apply ethics at each stage of the Machine Learning development pipeline, and to signal to researchers where further work is needed. The focus is exclusively on Machine Learning, but it is hoped that the results of this research may be easily applicable to other branches of AI. The article outlines the research method for creating this typology, the initial findings, and provides a summary of future research needs.

**Introduction**

As the availability of data on almost every aspect of life, and the sophistication of machine learning (ML) techniques, has increased (Lepri, Oliver, Letouzé, Pentland, & Vinck, 2018) so have the opportunities for improving both public and private life (Floridi & Taddeo, 2016). Society has greater control than it has ever had over outcomes related to: (1) who people can become; (2) what people can do; (3) what people can achieve; and (4) how people can interact with the world (Floridi et al., 2018,). However, growing concerns about the ethical challenges posed by the increased use of ML in particular, and Artificial Intelligence (AI) more generally, threaten to put a halt to the advancement of beneficial applications, unless handled properly.

Balancing the tension between supporting innovation, so that society's right to benefit from science is protected (Knoppers & Thorogood, 2017), and limiting the potential harms associated with poorly-designed AI (and specifically ML in this context), (summarised in figure 1) is challenging. ML algorithms are powerful socio-technical constructs (Ananny & Crawford, 2018), which raise concerns that are as much (if not more) about people as they are about code (see Table1) (Crawford & Calo, 2016). Enabling the so-called dual advantage of 'ethical ML'—so that the opportunities are capitalised on, whilst the harms are foreseen and minimised or prevented ( Floridi et al., 2018) —requires asking difficult questions about design, development, deployment, practices, uses and users, as well as the data that fuel the whole life-cycle of algorithms (Cath, Zimmer, Lomborg, & Zevenbergen, 2018). Lessig was right all along: code is both our greatest threat and our greatest promise (Lessig & Lessig, 2006).

| Ethical Concern | Explanation |
|---|---|
| Inconclusive Evidence | Algorithmic conclusions are probabilities and therefore not infallible. This can lead to unjustified actions. For example, an algorithm used to assess credit worthiness could be accurate 99% of the time, but this would still mean that one out of a hundred applicants would be denied credit wrongly. |
| Inscrutable Evidence | A lack of interpretability and transparency can lead to algorithmic systems that are hard to control, monitor, and correct. This is the commonly cited 'black-box' issue. |
| Misguided Evidence | Conclusions can only be as reliable (but also as neutral) as the data they are based on, and this can lead to bias. For example, (Dressel & Farid, 2018) found that the COMPAS recidivism algorithm commonly used in pretrial, parole, and sentencing decisions in the United States, is no more accurate or fair than predictions made by people with little or no criminal justice expertise. |
| Unfair outcomes | An action could be found to be discriminatory if it has a disproportionate impact on one group of people. For instance, (Selbst, 2017) articulates how the adoption of predictive policing tools is leading to more people of colour being arrested, jailed or physically harmed by police. |
| Transformative effects | Algorithmic activities, like profiling, can lead to challenges for autonomy and informational privacy. For example, (Polykalas & Prezerakos, 2019) examined the level of access required to personal data by more than 1000 apps listed in the 'most popular' free and paid for categories on the Google Play Store. They found that free apps requested significantly more data than paid-for apps, suggested that the business model of these 'free' apps is the exploitation of the personal data. |
| Traceability | It is hard to assign responsibility to algorithmic harms and this can lead to issues with moral responsibility. For example, it may be unclear who (or indeed what) is responsible for autonomous car fatalities. An in depth ethical analysis of this specific issue is provided by(Hevelke & Nida-Rümelin, 2015) |

*Table 1: Ethical concerns related to algorithmic use based on the 'map' created by (Mittelstadt, Allo, Taddeo, Wachter, & Floridi, 2016)*

Rising to the challenge of designing 'ethical ML' is both essential and possible. Indeed those that claim that it is impossible are falling foul of the is-ism fallacy where they confuse the way things are with the way things can be (Lessig & Lessig, 2006), or indeed should be. It is possible to design an algorithmically-enhanced society pro-ethically[1] (Floridi, 2016b), so that it protects the values, principles, and ethics that society thinks are fundamental (Floridi, 2018). This is the message that social scientists, ethicists, philosophers, policymakers, technologists, and civil society have been delivering in a collective call for the development of appropriate governance mechanisms (D'Agostino & Durante, 2018) that will enable society to capitalise on the opportunities, whilst ensuring that human rights are respected (Floridi & Taddeo, 2016), and fair and ethical decision-making is maintained (Lipton, 2016).

---

[1] The difference between ethics by design and pro-ethical design is the following: *ethics by design* can be paternalistic in ways that constrain the choices of agents, because it makes some options less easily available or not at all; instead, *pro-ethical design* still forces agents to make choices, but this time the nudge is less paternalistic because it does not preclude a course of action but requires agents to make up their mind about it. A simple example can clarify the difference. A speed camera is a form of nudging (drivers should respect the speed limits) but it is pro-ethical insofar as it leaves to the drivers the freedom to choose to pay a ticket, for example in case of an emergency. On the contrary, in terms of ethics by design, speed bumps are a different kind of traffic calming measure designed to slow down vehicles and improve safety. They may seem like a good idea, but they involve a physical alteration of the road, which is permanent and leaves no real choice to the driver. This means that emergency vehicles, such as a medical ambulance, a police car, or a fire engine, must also slow down, even when responding to an emergency.

The purpose of the following pages is to highlight the part that technologists, or ML developers, can have in this broader conversation. Specifically, section 'Moving from Principles to Practice' discusses how efforts to data have been too focused on the 'what' of ethical AI (i.e. debates about principles and codes of conduct) and not enough on the 'how' of applied ethics. The 'Methodology' section outlines the research planned to contribute to closing this gap between principles and practice, through the creation of an 'applied ethical AI typology,' and the methodology for its creation. Section 'Framing the results,' provides the theoretical framework for interpreting the results. The 'Discussion of initial results' section summarises what the typology shows about the uncertain utility of the tools and methods identified as well as their uneven distribution. The section on 'A way forward' argues that there is a need for a more coordinated effort, from multi-disciplinary researchers, innovators, policymakers, citizens, developers and designers, to create and evaluate new tools and methodologies, in order to ensure that there is a 'how' for every 'what' at each stage of the Machine Learning pipeline. The penultimate section lists some of the limitations of this study. Finally, the last section, concludes that the suggested recommendations will be challenging to achieve, but it would be imprudent not to try.

**Moving from Principles to Practices**

On 22nd May 2019, the Organisation for Economic Co-operation and Development (OECD) announced that its thirty-six member countries, along with an additional six (Argentine, Brazil, Columbia, Costa Rica, Peru, and Romania), had formally agreed to adopt, what the OECD claims to be the first intergovernmental standard on Artificial Intelligence (AI) (OECD, 2019). Designed to ensure AI systems are robust, safe, fair and trustworthy, the standard consists of five complementary value-based principles, and five implementable recommendations to policymakers.

The values and recommendations are not new. Indeed, the OECD's *Recommendation of the Council on Artificial Intelligence* (OECD, n.d.) is only the latest among a list of more than 70 documents, published in the last three years, which make recommendations about the principles of the ethics of AI (Spielkamp et al., 2019; Winfield, 2019). This list includes documents produced by industry (Google[2], IBM[3], Microsoft[4], Intel[5]), Government (Montreal Declaration[6], Lords Select Committee[7], European Commission's High-Level Expert Group[8]), and academia (Future of Life Institute[9], IEEE[10], AI4People[11]). The hope of the authors of these documents is that the      principles put forward, can, as abstractions (Anderson & Anderson, 2018), act as normative constraints (Turilli, 2007) on the 'do's' and 'don'ts' of algorithmic use in society.

As (Jobin, Ienca, & Vayena, 2019) and (Floridi, 2019c) point out, this intense interest from such a broad range of stakeholders reflects not only the need for ethical guidance, but also the desire of those different parties to shape the 'ethical AI' conversation around their own priorities. This is an issue that is not unique to debates about the components of ethical ML, but something that the international human rights community has grappled with for decades, as disagreements over what they are, how many there are, what they are for, as well as what duties they impose on whom, and which values of human interests they are supposed to protect (Arvan, 2014), have never been resolved. It is significant, therefore, that there seems to be an emerging consensus amongst the members of the ethical ML community with regards to *what* exactly ethical ML should aspire to be.

A review of 84 ethical AI documents by (Jobin et al., 2019) fund that although no single principle featured in all of them, the themes of transparency, justice and fairness, non-maleficence, responsibility and privacy appeared in over half. Similarly, a systematic review of the literature on ethical technology revealed that the themes of privacy, security, autonomy, justice, human dignity, control of technology and the balance of powers, were recurrent (Royakkers, Timmer, Kool, & van Est, 2018). As argued by (Floridi & Cowls, 2019), taken together these themes 'define' ethically-aligned ML as that which is (a) beneficial to, and respectful of, people and the environment (**beneficence**); (b) robust and secure (**non-maleficence**); (c)respectful of

---

human values (**autonomy**); (d) fair (**justice**); and (e) explainable, accountable and understandable (**explicability**). Given this emergent consensus in the literature, it is unsurprising that these are also the themes central to the OECD standard. What is perhaps more surprising s that this agreement around the basic principles that ethical ML should meet is no longer limited to Europe and the Western world. Just three days after the OECD publication, the Beijing Academy of Artificial Intelligence (BAAI), an organisation backed by the Chinese Ministry of Science and technology and the Beijing municipal government, released its fifteen AI principles for: (a) research and development; (b) use; and (c) the Governance of AI (Knight, 2019), which when read in full, bear remarkable similarity to the common framework (see Figure 2).

| AI4People (published November 2018) (Floridi et al., 2018) | Five principles key to any ethical framework for AI (published March 2019) (Floridi & Clement-Jones, 2019) | Ethics Guidelines for Trustworthy AI (Published April 2019) (European Commission, 2019) | Recommendation of the Council of Artificial Intelligence (Published May 2019) (OECD,2019.) | Beijing AI Principles for R&D (Published May 2019) ('Beijing AI Principles', 2019) |
|---|---|---|---|---|
| **Beneficence** | AI must be beneficial to humanity | Respect for human autonomy | Inclusive growth, sustainable development and well-being | **Do Good:** (covers the need for AI to promote human society and the environment) |
| **Non-Maleficence** | AI must not infringe on privacy or undermine security | Prevention of harm | Robustness, security and safety | **Be Responsible:** (covers the need for researchers to be aware of negative impacts and take steps to mitigate them)<br><br>**Control Risks:** (covers the need for developers to improve the robustness and reliability of systems to ensure data security and AI safety) |
| **Autonomy** | AI must protect and enhance our autonomy and ability to take decisions and choose between alternatives | | **Human-centred values** and fairness | **For Humanity:** (covers the need for AI to serve humanity by conforming to human values including freedom and autonomy) |
| **Justice** | AI must promote prosperity and | Fairness | Human-centred values **and fairness** | **Be Diverse and Inclusive:** (covers the need for AI to benefit as many people as possible)<br><br>**Be Ethical:** (covers the need to make the system as fair as possible, minimising discrimination and bias) |
| **Explicability** | AI systems must be understandable and explainable | Explicability | Transparency and Explainability<br><br>Accountability | **Be Ethical:** (covers the need for AI to be transparent, explainable and predictable) |

*Table 2: Comparison of ethical principles in recent publications demonstrating the emerging consensus of 'what' ethical AI should aspire to be[12]*

This fragile[13] consensus means that there is now the outline of a shared foundation upon which one can build, and that can be used as a benchmark to communicate expectations and evaluate

---

[12] For a more detailed comparison see (Floridi & Cowls, 2019; Hagendorff, 2019)

[13] We say fragile here as there are gaps across the different sets of principles and all use slightly different terminology, making it hard to guarantee that the exact same meaning is intended in all cases. Furthermore, as these principles have no legal grounding there is nothing to prevent any individual country (or indeed company) from suddenly choosing to adopt a different set for purposes of convenience or competitiveness

deliverables. Co-design in AI would be more difficult without this common framework. It is, therefore, a necessary building block in the creation of an environment that fosters ethical, responsible, and beneficial ML, especially as it also indicates the possibility of a time when the distractive risk of ethics shopping[14] (Floridi, 2019) will be lessened. Yet, challenges remain,

The availability of these 'agreed' principles supports but does not yet bring about actual change in the *design* of algorithmic systems ( Floridi, 2019a). As (Hagendorff, 2019) notes, almost all of the guidelines that have been produced to date suggest that technical solutions exist, but very few provide technical explanations. As a result, developers are becoming frustrated by how little help is offered by highly abstract principles when it comes to the 'day job' (Peters & Calvo, 2019). This is reflected in the fact that 79% of tech workers report that they would like practical resources to help them with ethical considerations (Miller & Coldicott, 2019). Without this more practical guidance, other risks such as 'ethics bluewashing'[15] and 'ethics shirking'[16] remain (Floridi, 2019b).

Such risks, associated with a lack of practical guidance on *how* to produce ethical ML, make it clear that the ethical ML community needs to embark on the second phase of AI ethics: translating between the '*what*' and the '*how.*' This is likely to be hard work. The gap between principles and practice is large, and widened by complexity, variability, subjectivity, and lack of standardisation, including variable interpretation of the 'components' of each of the ethical principles (Alshammari & Simpson, 2017). Yet, it is not impossible if the right questions are asked (Green, 2018; Wachter, Mittelstadt, & Floridi, 2017) and closer attention is payed to how the design process can influence (Kroll, 2018) whether an algorithm is more or less 'ethically-aligned.'

The sooner we start doing this, the better. If we do not take on the challenge and develop usable, interpretable and efficacious mechanisms (Abdul, Vermeulen, Wang, Lim, & Kankanhalli, 2018) for closing this gap, the lack of guidance may (a) result in the costs of ethical mistakes outweighing the benefits of ethical success (even a single critical 'AI' scandal could stifle innovation): (b) undermine public acceptance of algorithmic systems; (c) reduce adoption of algorithmic systems; and (d) ultimately create a scenario in which society incurs significant opportunity costs (Cookson, 2018). Thus, the aim of this research project is is to identify the methods and tools already available to help developers, engineers, and designers of ML reflect on and apply 'ethics' (Adamson, Havens, & Chatila, 2019) so that they may know not only what to do or not to do, but also how to do it, or avoid doing it (Alshammari & Simpson, 2017). We hope that the results of this research may be easily applicable to other branches of AI.

---

[14] "*Digital ethics shopping* is the malpractice of choosing, adapting, or revising ("mixing and matching") ethical principles, guidelines, codes, frameworks or other similar standards (especially but not only in the ethics of AI), from a variety of available offers, in order to retrofit some pre-existing behaviours (choices, processes, strategies etc.) and hence justify them *a posteriori,* instead of implementing or improving new behaviours by benchmarking them against public, ethical standards" (Floridi, 2019b)

[15] "*Ethics bluewashing* is the malpractice of making unsubstantiated or misleading claims about, or implementing superficial measures in favour of, the ethical values and benefits of digital processes, products, services, or other solutions in order to appear more digitally-ethical than one is." (Floridi, 2019b)

[16] "*Ethics shirking* is the malpractice of doing increasingly less "ethical work" (such as fulfilling duties, respecting rights, honouring commitments, etc.) in a given context the lower the return of such ethical work in that context is mistakenly perceived to be." (Floridi, 2019b)

**Methodology**

With the aim of identifying the methods and tools available to help developers, engineers and designers of ML reflect on and apply 'ethics' in mind, the first task was to design a typology, for the very practically minded ML community (Holzinger, 2018), that would 'match' the tools and methods identified to the ethical principles outlined in table 2 (summarised as beneficence, non-maleficence, autonomy, justice, and explicability).

To create this typology, and inspired by (Saltz & Dewar, 2019) who produced a framework that is meant to help data scientists consider ethical issues at each stage of a project, the ethical principles were combined with the stages of algorithmic development outlined in the overview of the Information Commissioner's Office (ICO) auditing framework for artificial intelligence and its core components[17], as shown in table 3. The intention is that this encourages ML developers to go between decision and ethical principles regularly.

|  | Business and use-case development<br>Problem/improvements are defined and use of AI is proposed | Design Phase<br>The business case is turned into design requirements for engineers | Training and test data procurement<br>Initial data sets are obtained to train and test the model | Building<br>AI application is built | Testing<br>The system is tested | Deployment<br>When the AI system goes live | Monitoring<br>Performance of he system is assessed |
|---|---|---|---|---|---|---|---|
| **Beneficence** |  |  |  |  |  |  |  |
| **Non-Maleficence** |  |  |  |  |  |  |  |
| **Autonomy** |  |  |  |  |  |  |  |
| **Justice** |  |  |  |  |  |  |  |
| **Explicability** |  |  |  |  |  |  |  |

*Table 3: 'Applied AI Ethics' Typology comprising ethical principles and the stages of algorithmic development*

The second task was to identify the tools and methods, and the companies or individuals researching and producing them, to fill the typology. There were a number of different ways this could have been done. For example, (Vakkuri, Kemell, Kultanen, Siponen, & Abrahamsson, 2019) sought to answer the question 'what practices, tools or methods, if any, do industry professionals utilise to implement ethics in to AI design and development?' by conducting interviews at five companies that develop AI systems in different fields. However, whilst analysis of the interviews revealed that the developers were aware of the potential importance of ethics in AI, the companies seemed to provide them with no tools or methods for implementing ethics. Based on a hypothesis that these findings did not imply the non-existence of applied-ethics tools and methods, but rather a lack of progress in the translation of available tools and methods from academic literature or early-stage development and research, to real-life use, this study used the traditional approach of providing an overarching assessment of a research topic, namely a literature review (Abdul et al., 2018).

---

[17] More detail is available here: https://ai-auditingframework.blogspot.com/2019/03/an-overview-of-auditing-framework-for_26.html

Scopus[18], arXiv[19] and PhilPapers[20], as well as Google search were searched. The Scopus, arXiv and Google Search searches were conducted using the terms outlined in table 4. The PhilPapers search was unstructured, given the nature of the platform, and instead the categories also shown in table 4 were reviewed. The original searches were run in February 2019, but weekly alerts were set for all searches and reviewed up until mid-July 2019. Every result (of which there were originally over 1,000) was checked for *relevance* – either in terms of theoretical framing or in terms of the use of the tool – *actionability* by ML developers, and *generalisability* across industry sectors. In total, 425 sources[21] were reviewed. They provide a practical or theoretical contribution to the answer of the question: 'how to develop an ethical algorithmic system.[22]'

| Scopus, Google and arXiv Search Terms (all searched with AND Machine Learning OR Artificial Intelligence) | Category of PhilPapers Reviewed |
|---|---|
| Ethics | Information Ethics |
| Public Perception | Technology Ethics |
| Intellectual Property | Computer Ethics |
| Business Model | Autonomy in Applied Ethics |
| Evaluation | Beneficence in Applied Ethics |
| Data Sharing | Harm in Applied Ethics |
| Impact Assessment | Justice in Applied Ethics |
| Privacy | Human Rights in Applied Ethics |
| Harm | Applied Ethics and Normative Ethics |
| Legislation | Responsibility in Applied Ethics |
| Regulation | Ethical Theories in Applied Ethics |
| Data Minimisation | |
| Transparency | |
| Bias | |
| Data Protection | |

*Table 4 showing the search terms used to search Scopes, arXiv and Google and the categories reviewed on PhilPapers*

---

[18] Scopus is the largest abstract and citation database of peer-reviewed literature: scientific journals, books and conference proceedings: https://www.scopus.com/home.uri
[19] arXiv provides open access to over 1,532,009 e-prints in the fields of physics, mathematics, computer science, quantitative biology, quantitative finance, statistics, electrical engineering and systems science, and economics: https://arxiv.org/
[20] PhilPapers is an index and bibliography of philosophy which collates research content from journals, books, open access archives and papers from relevant conferences such as IACAP. The index currently contains more than 2,377,536 entries. https://philpapers.org/
[21] This total includes references related specifically to discourse ethics after an anonymous reviewer made the excellent suggestion that this literature be used as a theoretical frame for the typology.
[22] The full list of sources can be accessed here: https://medium.com/@jessicamorley/applied-ai-ethics-reading-resource-list-ed9312499c0a

The third, and final task, was to review the recommendations, theories, methodologies, and tools outlined in the reviewed sources, and identify where they may fit in the typology. To do this, each of the high-level principles (beneficence, non-maleficence, autonomy, justice and explicability) were translated into tangible system requirements that reflect the meaning of the principles. This is the approach taken by the EU's High Level Ethics Group for AI and outlined in Chapter II of *Ethics Guidelines for Trustworthy AI: Realising Trustworthy AI* which "offers guidance on the implementation and realisation of Trustworthy AI, via a list of (seven) requirements that should be met, building on the principles" (p.35 European Commission, 2019).

This approach is also used in the disciplinary ethical guidance produce for internet-mediated researchers by the Belmont Report (Anabo, Elexpuru-Albizuri, & Villardón-Gallego, 2019), and by (La Fors, Custers, & Keymolen, 2019) who sought to integrate existing design-based ethical approaches for new technologies by matching lists of values the practical abstraction from mid-level ethics (principles) to what (Hagendorff, 2019) calls 'microethics.' This translation is a process that gradually reduces the indeterminacy of abstract norms to produce desiderata for a 'minimum-viable-ethical-(ML)product' (MVEP) that can be used by people who have various disciplinary backgrounds, interests and priorities (Jacobs & Huldtgren, 2018). The outcome of this translation process is shown in Figure 5.

| Principle | Beneficence | Non-Maleficence | Autonomy | Justice | Explicability |
|---|---|---|---|---|---|
| Requirements | **Stakeholder participation:** to develop systems that are trustworthy and support human flourishing, those who will be affected by the system should be consulted<br><br>**Protection of fundamental rights**<br><br>**Sustainable and environmentally friendly AI:** the system's supply chain should be assessed for resource usage and energy consumption<br><br>**Justification:** the purpose for building the system must be clear and linked to a clear benefit – system's should not be built for the sake of it. | **Resilience to attack and security:** AI systems should be protected against vulnerabilities that can allow them to be exploited by adversaries.<br><br>**Fallback plan and general safety:** AI systems should have safeguards that enable a fallback plan in case of problems.<br><br>**Accuracy:** for example, documentation that demonstrates evaluation of whether the system is properly classifying results.<br><br>**Privacy and Data Protection:** AI systems should guarantee privacy and data protection throughout a | **Human agency:** users should be able to make informed autonomous decisions regarding AI systems<br><br>**Human oversight:** may be achieved through governance mechanisms such as human-on-the-loop, human-in-the-loop, human-in-command. | **Avoidance of unfair bias**<br><br>**Accessibility and universal design**<br><br>**Society and democracy:** the impact of the system on institutions, democracy and society at large should be considered<br><br>**Auditability:** the enablement of the assessment of algorithms, data and design processes.<br><br>**Minimisation and reporting of negative impacts:** measures should be taken to identify, assess, document, minimise and respond to potential negative impacts of AI systems | **Traceability:** the data sets and the processes that yield the AI system's decision should be documented<br><br>**Explainability:** the ability to explain both the technical processes of an AI system and the related human decisions<br><br>**Interpretability** |

| | | system's entire lifecycle.<br><br>**Reliability and Reproducibility:** does the system work the same way in a variety of different scenarios.<br><br>**Quality and integrity of the data:** when data is gathered it may contain socially constructed biases, inaccuracies, errors and mistakes – this needs to be addressed<br><br>**Social Impact:** the effects of system's on people's physical and mental wellbeing should be carefully considered and monitored. | | **Trade-offs:** when trade-offs between requirements are necessary, a process should be put in place to explicitly acknowledge the trade-off, and evaluate it transparently<br><br>**Redress:** mechanism should be in place to respond when things go wrong. | |

*Table 5 showing the connection between high-level ethical principles and tangible system requirements as adapted from the methodology outlined in Chapter II of the European Commission's "Ethics Guidelines for Trustworthy AI"*

## Framing the results

The full typology is available here [ redacted for anonymity ]. The purpose of presenting it is not to imply that it is 'complete,' nor that the tools and methodologies highlighted are the best, or indeed the only, means of 'solving' each of the individual ethical problems. How to apply ethics to the development of ML is an open question that can be solved in a multitude of different ways at different scales and in different contexts (Floridi, 2019a). It would, for example, be entirely possible to complete the process using a different set of principles and requirements. Instead, the goal is to provide a synthesis of what tools are currently available to ML developers to encourage the progression of ethical AI from principles to practice and to signal clearly, to the 'ethical AI' community at large, where further work is needed.

Additionally, the purpose of presenting the typology is not to give the impression that the tools act as means of translating the principles into definitive 'rules' that technology developers should adhere to, or that developers must always complete one 'task' from each of the boxed. This only promotes ethics by 'tick-box' (Hagendorff, 2019). Instead, the typology is intended to eventually be an online searchable database so that developers can look for the appropriate tools and methodologies for their given context, and use them to enable a shift from a prescriptive 'ethics-by-design' approach to a dialogic, pro-ethical design approach (Anabo et al., 2019; Floridi, 2019a).

In this sense, the tools and methodologies represent a pragmatic version of Habermas's discourse ethics[23] (Mingers & Walsham, 2010). In his theory, Habermas (1983, 1991) argues that morals and norms are not 'set' in a top-down fashion but emerge from a process where those with opposing views, engage in a process where they rationally consider each other's arguments, give reasons for their position and, based upon the greater understanding that results, reassess their position until all parties involved reach a universally agreeable decision (Buhmann, Paßmann, & Fieseler, 2019). This is an approach commonly used in both business and operational research ethics, where questions of 'what *should* we do?' (as opposed to what *can* we do?) arise (Buhmann et al., 2019; Mingers, 2011). This is a rationalisation process that involves a fair consideration of the practical, the good and the just, and normally relies heavily on language (discussion), for both the emergence of agreed upon norms or standards, and their reproduction. In the present scenario of developers rationalising ML design decisions to ensure that they are ethically-optimised, the tools and methods in the typology replace the role of language and act as the medium for identifying, checking, creating and re-examining ideas and giving fair consideration to differing interests, values and norms (Heath, 2014; Yetim, 2019). For example, the data nutrition tool (Holland, Hosny, Newman, Joseph, & Chmielinski, 2018) provides a means of prompting a discussion and re-evaluation of the ethical implications of using a specific dataset for an ML development project, and the audit methodologies of (Diakopoulos, 2015) ensure that external voices, who may have an opposing view as to whether or not an ML-system in use is ethically-aligned, have a mechanism for questioning the rational of design decisions and requesting their change if necessary. It is within this frame that we present an overview of our findings in the next section.

**Discussion of initial results**

Interpretation of the results of the literature review and the resulting typology are likely to be context specific. Those with different disciplinary backgrounds (engineering, moral philosophy, sociology etc.) will see different patterns, and different meanings in these patterns. This kind of multidisciplinary reflection on what the presence or absence of different tools and methods, and their function, might mean is to be encouraged. To start the conversation, this section highlights the following three headings:

1. an overreliance on 'explicability';
2. a focus on the need to 'protect' the individual over the collective; and

---

[23] We would like to thank one of the anonymous reviewers for suggesting this framing, it represents a significant improvement to the theoretical grounding of this paper.

3. a lack of usability

They are interrelated, but for the sake of simplicity, let us analyse each separately.

*Explicability as the all-encompassing principle*

To start with the most obvious observation: the availability of tools and methods is not evenly distributed across the typology, either in terms of the ethical principles or in terms of the stages of development. For example, whilst a developer looking to ensure their ML algorithm is 'non-maleficent' has a section of tools available to them for each development stage – as highlighted in table 6 – the tools and methods designed to enable developers to meet the principle or 'beneficence' are almost all intended to be used during the initial planning stages of development (i.e. business and use-case development design phases). However, the most noticeable 'skew' is towards post-hoc 'explanations;' with those seeking to meet the principle of explicability during the testing phase having the greatest range of tools and methods from which to choose.

There are likely to be several reasons for this, but two stand out. The first and simpler is that the 'problem' of 'interpreting' an algorithmic decision seems tractable from a mathematical standpoint, so the principle of explicability has come to be seen as the most suitable for a technical fix (Hagendorff, 2019). The second is that 'explicability' is not, from a moral philosophy perspective, a moral principle like the other four principles. Instead, it can be seen as a second order principle, that has come to be of vital importance in the ethical-ML community because, to a certain extent, it is linked with all the other four principles[24]. Indeed, it is argued that if a system is explicable (explainable and interpretable) it is inherently more transparent and therefore more accountable in terms of its decision-making properties and the extent to which they include human oversight and are fair, robust and justifiable (Binns et al., 2018; Cath, 2018; Lipton, 2016).

Assuming temporarily that this is indeed the case[25], and that by dint of being explicable an ML system can more easily meet the principles of beneficence, non-maleficence, autonomy and justice, then the fact that the ethical ML community has focused so extensively on developing tools for 'explanations' may not seem problematic. However, as the majority of tools and methods that sit in the concentration at the intersection of explicability and testing are primarily statistical in nature, this would be a very mechanistic view because such 'solutions' - e.g. LIME (Ribeiro, Singh, & Guestrin, 2016), SHAP (Lundberg & Lee, 2017), Sensitivity Analysis (Oxborough et al., 2018) – do not really succeed in helping developers provide meaningful *explanations* (Edwards & Veale, 2018) that give individuals greater control over what is being *inferred* about them from their data. As such, the existence of these tools is at most necessary but not sufficient.

---

[24] We would like to thank one of the anonymous reviewers for making this important point.

[25] It is entirely possible that this is not always the case and that there may be instances where an explicable system has, for example, still had a negative impact on autonomy. Additionally, this view that transparency as explanation is key to accountability is one that is inherently western in perspective and those of other cultures may have a different viewpoint. We make the assumption here for simplicity's sake.

From a more humanistic, and realistic perspective, in order to satisfy all the five principles a system needs to be *designed* from the very beginning to be a transparent sociotechnical system (Anany & Crawford, 2018). To achieve this level of transparency, accountability or explicability, it is essential that those analysing a system are able to "understand what it was designed to do, how it was designed to do that, and why it was designed in that particular way instead of some other way" (Kroll, 2018, p4). This kind of scrutiny will only be possible through a combination of tools or processes that facilitate auditing, transparent development, education of the public, and social awareness of developers (Burrell, 2016). As such, there should ideally be tools and methods available for each of the boxes in the typology, accepting that there may be areas of the typology which are more significant for ML practitioners than others.

Furthermore, available of tools and methods in a variety of typology areas is also important in the context of culturally and contextually specific ML ethics. Not all of the principles will be of equal importance in all contexts. For example, in the case of national security systems non-maleficence may be of considerably higher importance than explicability. If the community prioritises the development of tools and methods for one of the principles over the others, it will be denying itself the opportunity for such flexibility.

| | Business and use-case development Problem/improvements are defined and use of AI is proposed | Design Phase The business case is turned into design requirements for engineers | Training and test data procurement Initial data sets are obtained to train and test the model | Building AI application is built | Testing The system is tested | Deployment When the AI system goes live | Monitoring Performance of the system is assessed |
|---|---|---|---|---|---|---|---|
| Beneficence | | | | | | | |
| Non-maleficence | (Cavoukian, Taylor, & Abrams, 2010) outline 7 foundational principles for Privacy by Design: 1.Proactive not reactive: preventative not reactive. 2.Privacy as the default | (Oetzel & Spiekermann, 2014) set out a step-by-step privacy impact assessment (PIA) to enable companies | (Antignac, Sands, & Schneider, 2016) provide the python code to create *DataMin* (a data minimiser – a pre- | (Kolter & Madry, 2018) provide a practical introduction, from a mathematical and coding perspective, to the topic of adversarial robustness | (Dennis, Fisher, Lincoln, Lisitsa, & Veres, 2016) outline a methodology for verifying the decision- | (*AI Now Institute Algorithmic Accountability Policy Toolkit*) provides a list of questions policy and legal advocates will want to ask when considering introducing an automated system into a public service and provides detailed guidance on where in the procurement process to | (Makri & Lambrinoudakis, 2015)outline a structured privacy audit procedure based on the most widely adopted privacy principles: |

| | | | | | | |
|---|---|---|---|---|---|---|
| 3. Privacy embedded into design 4. Full functionality = positive sum, not zero sum 5. End-to-end lifecycle protection 6. Visibility and Transparency 7. Respect for user privacy | to achieve 'privacy-by-design' | processor modifying the input of data to ensure only the data needed are available to the program) as a series of Java source code files which can be run on the data sources points before disclosing the data. | with the idea being that it is possible to train deep learning classifiers to be resistant to adversarial attacks: https://adversarial-ml-tutorial.org/ | making of an autonomous agent to confirm that the controlling agent never deliberately makes a choice it believes to be unsafe | ask questions about accountability and potential harm https://ainowinstitute.org/aap-toolkit.pdf | -Purpose specification -Collection limitation -Data quality -Use retention and disclosure limitation -Safety safeguards -Openness -Individual participation -Accountability |
| **Autonomy** | | | | | | |
| **Justice** | | | | | | |
| **Explicability** | | | | | | |

*Table 6: Applied AI ethics typology with illustrative non-maleficence example. A developer looking to ensure their ML solutions meets the principle of non-maleficence can start with the foundational principles of privacy by design (Cavoukian et al., 2010) to guide ideation appropriately, use techniques such as data minimisation (Antignac et al., 2016), training for adversarial robustness (Kolter & Madry, 2018), and decision-making verification (Dennis et al., 2016) in the train-build-test phases, and end by launching the system with an accompanying privacy audit procedure (Makri & Lambrinoudakis, 2015)*

*An individual focus*

The next observation of note is that few of the available tools surveyed provide meaningful ways to assess, and respond to, the impact that the data-processing involved in their ML algorithm has on an individual, and even less on the impact on society as a whole (Poursabzi-Sangdeh, Goldstein, Hofman, Vaughan, & Wallach, 2018). This is evident from the very sparsely populated 'deployment' column of the typology. Its emptiness implies that the need for pro-ethically designed human-computer interaction (at an individual level) or networks of ML systems (at a group level) has been paid little heed. This is likely because it is difficult to translate complex human behaviour into design tools that are simple to use and generalisable.

This might not seem particularly importance, but the impact this has on the overall acceptance of AI in society could be significant. For example, it is unlikely that counterfactual explanations[26] (i.e. if input variable $x$ had been different, the output variable $y$ would have been different as well) will do anything to improve the interpretability of recommendations made by black-box systems for the average member of the public or the technical community. If such methods become the *de facto* means of providing 'explanations,' the extent to which the 'algorithmic society' is interpretable to the general public will be very limited. And counterfactual

---

[26] See for example (Johansson, Shalit, & Sontag, 2016; Lakkaraju, Kleinberg, Leskovec, Ludwig, & Mullainathan, 2017; Russell, Kusner, Loftus, & Silva, 2017; Wachter, Mittelstadt, & Floridi, 2017)

explanations could easily be embraced by actors uninterested in providing factual explanations, because the counterfactual ones provide a vast menu of options, which may easily decrease the level of responsibility of the actor choosing it. For example, if a mortgage provider does not offer a mortgage, the factual reasons may be a bias, for example the gender of the applicant, but the provider could choose from a vast menu of innocuous, counterfactual explanations – if some variable $x$ had been different the mortgage might have been provided – e.g., a much higher income, more collaterals, lower amount, and so forth, without ever mentioning the factual cause, i.e. the gender of the applicant. All this could considerably limit the level of trust people are willing to place in such systems.

This potential threat to trust is further heightened by the fact that the lack of attention paid to impact means that ML developers are currently hampered in their ability to develop systems that promote user's (individual or group's) autonomy. For example, currently there is an assumption that prediction = decision, and little research has been done (in the context of ML) on how people translate predictions into actionable decisions. As such, tools that, for example, help developers pro-ethically design solutions that do not overly restrict the user's options in acting on this prediction (i.e. tools that promote the user's autonomy) are in short supply (Kleinberg, Lakkaraju, Leskovec, Ludwig, & Mullainathan, 2017). F users feel as though their decisions are being curtailed and controlled by systems that they do not understand, it is very unlikely that these systems will meet the condition of social acceptability, never mind the condition of social preferability which should be the aim for truly ethically designed ML (Floridi & Taddeo, 2016).

*A lack of usability*

Finally, the tools and methods included in the typology are positioned as discourse aids, designed to facilitate and document rational decisions about trade-offs in the design process that may make an ML system more or less ethically-aligned. It is possible to see the *potential* for the tools identified to play this role. For example, at the "beneficence → use-case → design" intersection, there are a number of tools highlighted to help elicit social values. These include the responsible research and innovation methodology employed by the European Commission's Human Brain Project (Stahl & Wright, 2018), the field guide to human-centred design (idea.org) and Involve and DeepMind's guidance on stimulating effective public engagement on the ethics of artificial intelligence (Involve & DeepMind). Such tools and methods could be used to help designers pro-ethically deal with value pluralism (i.e. variation in values across different population groups). However, the vast majority of these tools and methods are not actionable as they offer little help on how to use them in practice (Vakkuri et al., 2019). Even when there are open-source code libraries available, documentation is often limited, and the skill-level required for use is high.

This overarching lack of usability of the tools and methods highlighted in the typology means that, although they are promising, they require more work before being 'production-ready.' As a result, applying ethics still requires considerable amounts of effort on behalf of the ML developers undermining one of the main aims of developing and using technologically-based 'tools': to remove friction from applied ethics. Furthermore, until these tools are embedded in practice and tested in the 'real world,' it is extremely unclear what impact they will have on the overall 'governability' of the algorithmic ecosystem. For example, (Binns, 2018a) asks how an accountable system actually will be held accountable for an 'unfair' decision in a way that is acceptable to all. This makes it almost impossible to measure the impact, 'define success', and document the performance (Mitchell et al., 2019) of a new design methodology or tool. As a result, tehre is no clear problem statement (and therefore no clear business case) that the ML community can use to justify time and financial investment in developing much-needed tools and techniques that truly enable pro-ethical design. Consequently, there is no guaranatee that the so-called discursive devices do anything other than help the groups in society who already have the loudest voices embed and protect their values in design tools, and then into the resultant ML systems.

**A way forward**

Social scientists (Matzner, 2014) and political philosophers (from Rousseau and Kant, to Rawls and Habermas) (Binns, 2018b), are used to dealing with the kind of plurality and subjectivity informing the entire ethical ML field (Bibal & Frénay, 2016). Answering questions such as, what happens when individual level and group level 'ethics' interact, and what key terms such as 'fairness,' 'accountability,' 'transparency' and 'interpretability' actually mean when there are currently a myriad definitions (Ananny & Crawford, 2018; Bibal & Frénay, 2016; Doshi-Velez & Kim, 2017; Friedler, Scheidegger, & Venkatasubramanian, 2016; Guidotti et al., 2018; Kleinberg, Mullainathan, & Raghavan, 2016; Overdorf, Kulynych, Balsa, Troncoso, & Gürses, 2018; Turilli & Floridi, 2009) is standard fare for individuals with social science, economy, philosophy or legal training.  This is why (Nissenbaum, 2004) argues for a contextual account of privacy, one that recognises the varying nature of informational norms (Matzner, 2014) and (Kemper & Kolkman, 2018) state that transparency is only meaningful in the context of a defined critical audience.

The ML developer community, in contrast, may be less used to dealing with *this* kind of difficulty, and more used to scenarios where there is at lest a seemingly quantifiable relationship between input and output. As a result, the existing approaches to designing and programming ethical ML fail to resolve what (Arvan, 2018) terms the moral-semantic trilemma, as almost all tools and methods highlighted in the typology are either too semantically strict, too semantically flexible, or overly unpredictable (Arvan, 2018).

Bridging together multi-disciplinary researchers into the development process of pro-ethical design tools and methodologies will be essential. A multi-disciplinary approach will help the ethical ML community overcome obstacles concerning social complexity, embrace uncertainty, and accept that: (1) AI is built on assumptions; (2) human behaviour is complex; (3) algorithms can have unfair consequences; (4) algorithmic predictions can be hard to interpret (Vaughan & Wallach, 2016); (5) trade-offs are usually inevitable; and (6) positive, ethical features are open to progressive increase, that is an algorithm can be increasingly fair, and fairer than another algorithm or a previous version, but makes no sense to say that it is fair or unfair n absolute terms (compare this to the case of speed: it makes sense to say that an object is moving quickly, or that it is fast or faster than another, but not that it is fast). The resulting collaborations are likely to be highly beneficial for the development of applied ethical tools and methodologies for at least three reasons.

First, it will help ensure that the tools and methods developed do not only protect value-pluralism in silico (i.e. the pluralistic values of developers) but also in society. Embracing uncertainty and disciplinary diversity will naturally encourage ML experts to develop tools that facilitate more probing and open (i.e. philosophical) questions (Floridi, 2019b) that will lead to more nuanced and reasoned answers and hence decisions about why and when certain trade-offs, for example, between accuracy and interpretability (Goodman & Flaxman, 2017), are justified, based on factors such as proportionality to risk (Holm, 2019).

Second, it will encourage a more flexible and reflexive approach to applied ethics that is more in-keeping with the way ML systems are actually developed: it is not think and *then* code, but rather think *and* code. In other words, it will accelerate the move away from the 'move fast and break things' approach towards an approach of 'make haste slowly' (*festina lente*) (Floridi, 2019a).

Finally, it would also mitigate a significant risk – posed by the current sporadic application of ethical-design tools and/or methods during different development stages – of the ethical principles having been written into the business and use-case, but coded out by the time a system gets to deployment.

To enable developers to embrace this vulnerable uncertainty, it will be important to promote the development of tools, like DotEveryone's agile consequence scanning event (DotEveryone, 2019), and the Responsible Double Diamond 'R2D2' (Peters & Calvo, 2019) that prompt developers to reflect on the impacts (both direct and indirect) of the solutions they are developing on the 'end user', and on how these impacts can be altered by seemingly minor design decisions at each stage of development. In other words, ML developers should regularly:

a) look back and ask: 'if I was abiding by ethical principles $x$ in my design *then,* am I still *now?* (as encouraged by *Wellcome Data Lab's* agile methodology (Mikhailov, 2019); and

b) look forward and ask: 'if I am abiding by ethical principles $x$ in my design *now,* should I continue to do so? And how? By using foresight methodologies (Floridi & Strait, Forthcoming; Taddeo & Floridi, 2018), such as *AI Now's* Algorithmic Impact Assessment Framework (Reisman, Schultz, Crawford, & Whittaker, 2018).

Taking this approach recognises that, in a digital context, ethical principles are not simply either applied or not, but regularly re-applied or applied differently, or better, or ignored as algorithmic systems are developed, deployed, configured (Ananny & Crawford, 2018) tested, revised and re-tuned (Arnold & Scheutz, 2018).

This approach to applied ML ethics of regular reflection and application will heavily rely on (i) the creation of more tools – especially to fill the white spaces of the typology (for the reasons discussed in the previous section) and (ii) acceleration of tools maturity level from research labs into production environments. To achieve (i)-(ii), society needs to come together in communities comprised of multi-disciplinary researchers (Cath, Wachter, Mittelstadt, Taddeo, & Floridi, 2017), including innovators, policymakers, citizens, developers and designers (Taddeo & Floridi, 2018), to foster the development of: (1) common knowledge and understanding; and (2) a common goal to be achieved from the development of tools and methodologies for applied AI ethics (Durante, 2010). These outputs will provide a reason, a mechanism, and a consensus to coordinate the efforts behind tool development. Ultimately, this will produce better results than the current approach, which allows a 'thousand flowers to bloom' but fails to create tools that fill int eh gaps (this is a typical 'intellectual market' failure), and may encourage competition to produce preferable options. The opportunity that this presents is too great to be delayed, the ML research community should start collaborating now with a specific focus on:

1. the development of a common language;

2. the creation of tools that ensure *people,* as individuals, groups and societies, are given an equal and meaningful opportunity to participate in the design of algorithmic solutions at each stage of development;

3. the evaluation of the tools that are currently in existence so that what works, what can be improved, and what needs to be developed can be identified;

4. a commitment to reproducibility, openness, and sharing of knowledge and technical solutions (e.g. software), also in view of satisfying (2) and supporting (3);

5. the creation of 'worked examples' of how tools have been used to satisfy one of the principles at each stage of the development and how consistency was maintained throughout the use of different tools'

6. the evaluation and creation of pro-ethical business models and incentive structures that balance the costs and rewards of investing in ethical AI across society, also in view of supporting (2)-(4).

**Limitations**

All research projects have their limitations and this one is no exception. The first is that the research question 'what tools and methods are available for ML developers to 'apply' ethics to each stage of the ML system design' is very broad. This lack of specificity meant that the available literature was excessive and growing all the time, making compromises from the perspective of practically essential. It is certain that such compromises, for example which databases to search and the decision to restrict the tools reviewed to those that were not industry sector-specific, have resulted in us missing a large number of tools and methods that are publicly available. Building on this, it is again, very likely that there are a number of proprietary applied ethics tools and methods being developed by private companies for internal or consulting purposes that we will have missed, for example the 'suite of customisable frameworks, tools and processes' that make up consulting firm PWC's "Responsible AI Toolkit" (PWC, 2019).

The second limitation is related to the design of the typology itself. As (La Fors et al., 2019) attest, the "neat theoretical distinction between different stages of technological innovation does not always exist in practice, especially not in the development of big data technologies." This implies that by categorising the tools by stage of development, we might be reducing their usability as developers in different contexts might follow a different pattern or feel as though it is 'too late' to, for example, engage in stakeholder engagement if they have reached the 'build' phase of their project, whereas the reality it is never too late.

Finally, the last limitations was already mentioned and concerns the lack of clarity regarding how the tools and methods that have been identified will improve the governability of algorithmic systems. Exactly *how* to govern ML remains an open question, although it appears that there is a growing acceptance among tech workers (in the UK at least) that government regulation will be necessary (Miller & Coldicott, 2019). The typology can at least be seen as a mechanism for facilitating co-regulation. Governments are increasingly setting standards and system requirements for ethical ML, but delegating the means for meeting these to the developers themselves (Clarke, 2019) – the tools and methods of the typology can be seen as the means of providing evidence of compliance. In this way, the typology (and the tools and methods it contains within) help developers take responsibility for embedding ethics in the part of the development, deployment, and use of ML solutions that they control (Coeckelbergh, 2012). The extent to which this makes a difference is yet to be determined.

## Conclusion

The realisation that there is a need to embed ethical considerations into the design of computational, specifically algorithmic, artefacts is not new. Both Alan Turing and Norbert Weiner were vocal about this in the 1940s and 1960s (Turilli, 2008). However, as the complexity of algorithmic systems and our reliance on them increases (Cath et al., 2017), so too does the need to be critical (Floridi, 2016a) AI governance (Cath, 2018) and design solutions. It is possible to design things to be better (Floridi, 2017), but this will require more coordinated and sophisticated approaches (Allen, Varner, & Zinser, 2000) to translating ethical principles into design protocols (Turilli, 2007).

This call for increased coordination is necessary. The research has shown that there is an uneven distribution of effort across the 'Applied AI Ethics' typology. Furthermore, many of the tools included are relatively immature. This makes it difficult to assess the scope of their use (resulting in Arvan's 2018 'moral-semantic trilemma') and consequently hard to encourage heir adoption by the practically-minded ML developers, especially when the competitive advantage of more ethically-aligned AI is not yet clear. Taking the time to complete any of the 'exercises' suggested by the methods reviewed, and investing in the development of new tools or methods that 'complete the pipeline', add additional work and costs to the research and development process. Such overheads may directly conflict with short-term, commercial incentives. Indeed, a full ethical approach to AI design, development, deployment, and use may represent a competitive disadvantage for any single 'first mover'. The threat that this short-termism poses to the development of truly ethical ML is significant. Unless a longer-term and sector-wide perspective in terms of return on investment can be encouraged – so that mechanisms are developed to close the gap between *what* and *how* – the lack of guidance may (a) result in the costs of ethical mistakes outweighing the benefits of ethical successes; (b) undermine public acceptance of algorithmic systems, even to the point of a backlash (Cookson, 2018); and (c) reduce adoption of algorithmic systems. Such a resultant lack of adoption could then turn into a loss of confidence from investors and research funders, and undermine AI research. Lack of incentives to develop AI ethically could turn into lack of interest in developing AI *tout court*. This would not be unprecedented. One only needs to recall the dramatic reduction in funding available for AI research following the 1973 publication of *Artificial Intelligence: A General Survey* (Lighthill, 1973) and its criticism of the fact that AI research had not lived up to its over-hyped expectations.

It this were to happen today, the opportunity costs that would be incurred by society would be significant (Cookson, 2018). The need for 'AI Ethics' has arisen from the fact that poorly designed AI systems can cause very significant harm. For example, predictive policing tools may lead to more people of colour being arrested, jailed or physically harmed by policy (Selbst, 2017). Likewise, the potential benefits of pro-ethically designed AI systems are considerable. This is especially true in the field of AI for Social Good where various AI applications are making possible socially good outcomes that were once less easily achievable, unfeasible, or unaffordable (Cowls, King, Taddeo, & Floridi, 2019). So, there is an urgent need to progress research in this area.

Constructive patience needs to be exercised, by society and by the ethical AI community, because such progress on the question of 'how' to meet the 'what' will not be quick, and there will definitely be mistakes along the way. The ML research community will have to accept this, trust that everyone is trying to meet the same end-goal, but also accept that it is unacceptable to delay any full commitment, when it is known how serious the consequences of doing nothing are. Only by accepting this can society by positive about the opportunities presented by AI to be seized, whilst remaining mindful of the potential costs to be avoided (Floridi et al., 2018).